# Carrier transport properties of the Group-IV ferromagnetic semiconductor $Ge_{1-x}Fe_x$ with and without boron doping

Yoshisuke Ban[a)], Yuki Wakabayashi, Ryota Akiyama[*], Ryosho Nakane, and Masaaki Tanaka[b)]

*Department of Electronic Engineering and Information Systems, The University of Tokyo, 7-3-1 Hongo, Bunkyo-ku, Tokyo 113-8656, Japan*

We have investigated the transport and magnetic properties of group-IV ferromagnetic semiconductor $Ge_{1-x}Fe_x$ films ($x$ = 1.0 and 2.3 %) with and without boron doping grown by molecular beam epitaxy (MBE). In order to accurately measure the transport properties of 100-nm-thick $Ge_{1-x}Fe_x$ films, (001)-oriented silicon-on-insulator (SOI) wafers with an ultra-thin Si body layer (~5 nm) were used as substrates. Owing to the low Fe content, the hole concentration and mobility in the $Ge_{1-x}Fe_x$ films were exactly estimated by Hall measurements because the anomalous Hall effect in these films was found to be negligibly small. By boron doping, we increased the hole concentration in $Ge_{1-x}Fe_x$ from ~$10^{18}$ cm$^{-3}$ to ~$10^{20}$ cm$^{-3}$ ($x$ = 1.0%) and to ~$10^{19}$ cm$^{-3}$ ($x$ = 2.3%), but no correlation was observed between the hole concentration and magnetic properties. This result presents a contrast to the hole-induced ferromagnetism in III-V ferromagnetic semiconductors.

a) Electronic mail: ban@cryst.t.u-tokyo.ac.jp
b) Electronic mail: masaaki@ee.t.u-tokyo.ac.jp
* Present address: *Institute of Materials Science, University of Tsukuba*

Ferromagnetic semiconductors (FMSs) have generated much interest from the viewpoint of physics, materials science, and possible applications to semiconductor-based spintronic devices. The origin of the ferromagnetism in FMSs was investigated by characterizing the magnetic properties and their correlation to other material properties, such as crystalline structure, carrier concentration, electrical properties, and electronic band structure. In III-V-based FMSs,[1,2] such as GaMnAs, it is well recognized that the ferromagnetism is induced by itinerant holes,[3] however, the origin of the ferromagnetism is still under debate[4] despite the tremendous research efforts for the past two decades. Also, there have been a number of studies on group-IV FMSs, especially on $Ge_{1-x}Mn_x$.[5-14] So far, the ferromagnetism in most of the $Ge_{1-x}Mn_x$ films is considered to come from intermetallic or Mn-rich metallic phases contained in the $Ge_{1-x}Mn_x$ films, such as nanocolumnar precipitations.[6,8]

Recently, we have grown epitaxial $Ge_{1-x}Fe_x$ films ($x$ = 2.0 – 17.5 %) and have shown that the $Ge_{1-x}Fe_x$ films have a diamond-type crystal structure without any other crystallographic phase of precipitates, and their ferromagnetic band structures characterized by magnetic circular dichroism (MCD) spectra were of diamond-type, identical with those of Ge bulk materials[15,16]. These results indicate that the $Ge_{1-x}Fe_x$ films are single-crystalline films with a single-ferromagnetic phase.[15,16] However, so far there is little information which helps to understand the origin of the ferromagnetism in group-IV FMSs. In particular, the transport properties of $Ge_{1-x}Fe_x$ films, including the carrier concentration and mobility, have not been investigated, although these are important characteristics to clarify the origin of ferromagnetism. Indeed, it was difficult to estimate the carrier concentrations in the $Ge_{1-x}Fe_x$ films by Hall effect measurements, because the anomalous Hall effect is dominant and the Hall resistances showed hysteresis due to the anomalous Hall effect when the Fe content $x$ was high ~ 10 %.[18] Studying the hole concentration dependence of the magnetic properties in $Ge_{1-x}Fe_x$ films is expected to be helpful to understand the ferromagnetism; if the ferromagnetism originates from itinerant holes in $Ge_{1-x}Fe_x$ (namely, carrier-induced ferromagnetism), ferromagnetic behavior is significantly enhanced with increasing the hole concentration.

In this paper, we study the transport and magnetic properties of $Ge_{1-x}Fe_x$ films with low Fe content ($x$ = 1.0 % and 2.3 %) where we can exactly estimate the hole concentration and mobility by Hall measurements. Owing to the low Fe content, the anomalous Hall effect was found to be negligibly small which enables us to estimate these parameters in the $Ge_{1-x}Fe_x$ films.  The hole concentration in the $Ge_{1-x}Fe_x$ films was varied by doping boron (B) atoms as acceptors.

We used (001)-oriented silicon-on-insulator (SOI) wafers as substrates, where the thickness of the Si layer was ~5 nm, in order to accurately measure the transport properties of $Ge_{1-x}Fe_x$ films by excluding the parallel conduction in the substrate.  After chemical etching of a thermally-oxidized $SiO_2$ top layer with HF, a substrate was installed into our molecular beam epitaxy (MBE) chamber equipped with a reflection high-energy electron diffraction (RHEED) system.  After thermal cleaning at a substrate temperature $T_S$ = 760°C for 30 sec, a 100-nm-thick $Ge_{1-x}Fe_x$ film was epitaxially grown at $T_S$ = 200°C with a rate of 120 nm/hour, where the Fe content $x$ was varied; 1.0, 2.3, 6.5, 10.5, and 14.0 %.  In some samples, a boron flux was also supplied to change the hole concentration during the MBE growth.  The RHEED pattern of $Ge_{1-x}Fe_x$ showed bright 2×2 streaks indicating diamond crystal structure with an atomically flat surface.  Finally, a 2-nm-thick Ge capping layer was grown on the $Ge_{1-x}Fe_x$ film.  Figure 1(a) shows the schematic structure and Table I lists the parameters of the samples.  Magnetization versus magnetic field ($M$-$H$) characteristics were measured using a Quantum Design MPMS superconducting quantum interference device (SQUID) magnetometer at 5 K.  The Curie temperature $T_C$ was estimated by the Arrott plots of MCD spectra measured at various temperatures (5 – 120 K).[19]  Using photolithography and wet etching with $H_2O_2$, the $Ge_{1-x}Fe_x$ films were fabricated into Hall bar-shaped devices with channel length of 200 μm and width of 50 μm.  The boron concentration $y$ was estimated by secondary ion mass spectroscopy (SIMS), and the hole concentration $p$ was estimated by the Hall effect measured at 300 K.  As a reference, a Ge film with a boron concentration of $4.4\times10^{19}$ $cm^{-3}$ was also grown on a SOI (001) substrate with the same structure, and a part of it was fabricated into a Hall bar-shaped device in the same manner.  Hereafter, this Ge film is referred to as boron-doped Ge film.

Figure 1(b) shows X-ray diffraction (XRD) $\theta$-$2\theta$ spectra of the samples, in which the assigned planes are indicated.  The spectra clearly indicate that all the samples have the $Ge_{1-x}Fe_x$ films with diamond crystal structure, and that other crystalline phases of Ge-Fe compounds were not observed.  This result is consistent with the characterizations using transmission electron microscopy and electron dispersive X-ray (EDX) spectroscopy in our previous studies.[16]  The lattice constant $a$ estimated from the $Ge_{1-x}Fe_x$ (004) peak in Fig. 1(b) decreased with increasing $x$, which is also consistent with the previous study.[18]  It was found that $a$ and the full width at half maximum (FWHM) of the $Ge_{1-x}Fe_x$ (004) peak were not changed by the boron doping when $y = 4.4\times10^{19}$ cm$^{-3}$. On the other hand, $a$ and the FWHM of the Ge$_{1-x}$Fe$_x$ (004) peak were changed by the boron doping when $y = 4.8\times10^{20}$ cm$^{-3}$; $a$ becomes smaller and FWHM is significantly larger by the boron doping, as shown in Table I. In other words, the boron doping of $y = 4.4\times10^{19}$ cm$^{-3}$ does not affect the lattice parameters of Ge$_{1-x}$Fe$_x$, but when $y = 4.8\times10^{20}$ cm$^{-3}$, a variation in the lattice parameters (from $y = 0$) is not negligible.

The resistivity was measured for the Hall bar-shaped devices using a four-terminal method. All the samples exhibited linear current-voltage characteristics in the temperature range from 5 K to 300 K (not shown here). Figure 2 (a) and (b) show the resistivity of the Ge$_{1-x}$Fe$_x$ films ((a) $x = 1.0$ %, (b) $x = 2.3$ %) as a function of temperature, where the yellow, orange, and red curves are the results of the Ge$_{1-x}$Fe$_x$ films ($x = 1.0$ %) with $y = 0$ (undoped, sample A), $y = 4.4\times10^{19}$ cm$^{-3}$ (sample B), and $y = 4.8\times10^{20}$ cm$^{-3}$ (sample C), respectively, and the green and blue curves are the results of the Ge$_{1-x}$Fe$_x$ films ($x = 2.3$ %) with $y = 0$ (undoped, sample D) and $y = 4.4\times10^{19}$ cm$^{-3}$ (sample E). In the figures, the resistivity of the boron-doped Ge film is also shown as a reference. The resistivity of the undoped ($y = 0$) Ge$_{1-x}$Fe$_x$ films (sample A and D) increases with decreasing temperature (below ~50 K),[18] whereas the resistivity of the boron-doped Ge$_{1-x}$Fe$_x$ films (sample B, C, and E) is almost constant in whole the temperature range just like the boron-doped Ge film. Thus, we obtained the metallic Ge$_{1-x}$Fe$_x$ films ($x = 1.0$ % and 2.3 %) by boron doping, which have not been obtained in the previous studies. On the other hand, we found that the resistivity of the boron-doped Ge$_{1-x}$Fe$_x$ films ($x = 6.5, 10.5,$ and 14.0 %) with $y = 4.4\times10^{19}$ cm$^{-3}$ increases with decreasing temperature (not shown here), meaning that these samples are not metallic. From these results, the activation rate of boron is significantly low when the Fe content exceeds 2.3 %. Since the purpose of this study is to investigate the effect of hole concentration on the properties of Ge$_{1-x}$Fe$_x$ films, hereafter we focus on the Ge$_{1-x}$Fe$_x$ films with $x = 1.0$ and 2.3 %.

To estimate the hole concentration $p$ in the temperature range of 5 – 300 K, the Hall voltage was measured for the devices with a constant current in the range of 10 μA to 1 mA while a perpendicular magnetic field was swept from -1 to 1 T. In general, the Hall coefficient $R_H$ of magnetic materials consists of the ordinary Hall effect and the anomalous Hall effect. The anomalous Hall effect which is proportional to the magnetization decreases with increasing temperature and becomes negligible well above the Curie temperature. In the boron-doped Ge$_{1-x}$Fe$_x$ films ($x = 1.0$ and 2.3 %; sample B, C, and E), Hall voltage versus magnetic field characteristics are linear without hysteresis and $R_H$ was almost constant in the whole temperature range (5 – 300 K). Thus, these temperature-insensitive $R_H$ data are caused by the ordinary Hall effect, since the contribution of the anomalous Hall effect, which is temperature-sensitive, is negligible. This is also supported by the result that the resistivity obtained for the boron-doped $Ge_{1-x}Fe_x$ films (sample B, C, and E) are temperature insensitive, as shown in Fig. 2 (a) and (b). Consequently, the hole concentration $p$ of the $Ge_{1-x}Fe_x$ films ($x$ = 1.0 and 2.3 %) can be estimated from $R_H = 1/qp$, where $q$ is the elementary charge. Figure 2 (c) and (d) show $p$ versus temperature characteristics obtained for the boron-doped $Ge_{1-x}Fe_x$ ((c) $x$ = 1.0 % and (d) $x$ = 2.3 %) films with $y$ = 4.4×10$^{19}$ and 4.8×10$^{20}$ cm$^{-3}$, in which the data of the boron-doped Ge film is also shown for comparison. As shown in Fig. 2 (c) and (d), we estimated the exact hole concentrations $p$ of $Ge_{1-x}Fe_x$, and succeeded in controlling $p$ in the range of 7×10$^{17}$ to 2×10$^{20}$ cm$^{-3}$. The activation rates of boron estimated by $p/y$, where $y$ is the boron concentration measured by SIMS, are 63 % for sample B and 39 % for sample E. In Fig. 2 (c), since the $p$ values of the boron-doped Ge film ($y$ = 4.4×10$^{19}$ cm$^{-3}$) and sample B ($x$ = 1.0 %, $y$ = 4.4×10$^{19}$ cm$^{-3}$) are same, the activation rate of boron in both the films are identical. On the other hand, in Fig. 2 (d), since the $p$ value of sample E ($x$ = 2.3 %, $y$ = 4.4×10$^{19}$ cm$^{-3}$) is slightly smaller than that of the boron-doped Ge film, the activation rate of boron was decreased due to the relatively higher Fe composition of $x$ = 2.3 %.

We estimated the hole mobility $\mu$ from $1/\rho = qp\mu$, where $p$ and the resistivity $\rho$ shown in Fig. 2 (a) − (d) were used. Figure 2 (e) and (f) show the temperature dependence of $\mu$ of the boron-doped $Ge_{1-x}Fe_x$ films ((e) $x$ = 1.0 %, (f) $x$ = 2.3 %) and the boron-doped Ge film. In the $Ge_{1-x}Fe_x$ films with $x$ = 1.0 %, $\mu$ at 300 K was 56 cm$^2$/sV for sample B ($y$ = 4.4×10$^{19}$ cm$^{-3}$), 43 cm$^2$/sV for sample C ($y$ = 4.8×10$^{20}$ cm$^{-3}$), and 46 cm$^2$/sV for sample A ($y$ = 0), respectively. These values are the same order of magnitude as typical values of the hole mobility in boron-doped Ge films,[17] although the doping concentration of Fe (1.0 %) is higher than a typical acceptor or donor doping concentration in semiconductors. This suggests that Fe atoms are not ionized but neutral. In Fig. 2 (e) and (f), $\mu$ of the $Ge_{1-x}Fe_x$ film decreased with increasing $x$ when $y$ is same, and these results are reasonable because the mobility usually decreases with increasing the impurity density in semiconductors.[17]

On the other hand, $\mu$ of the $Ge_{1-x}Fe_x$ films without boron doping ($y$ = 0; sample A and D) is greatly reduced at low temperatures as shown in Fig. 2 (e) and (f), which reflects the large increase in the resistivity of these films at low temperatures as shown in Fig. 2 (a) and (b). This result suggests that the conduction is dominated by hopping of holes in these undoped samples since the temperature dependence of $\mu$ cannot be fitted by neither phonon scattering ($\mu \propto T^{-3/2}$) nor by ionized impurity scattering ($\mu \propto T^{3/2}$). A possible mechanism of the hole conduction is hopping of holes via neutral impurity Fe atoms. We think that most of the Fe atoms are neutral, as suggested by the relatively high mobility data in Fig. 2 (e) and (f) and by the very low activation rate of Fe (less than 0.5 %) estimated by the data in Fig. 2 (c) and (d).

Figure 3 shows *M-H* curves measured with a perpendicular magnetic field for the boron-doped $Ge_{1-x}Fe_x$ films ($x$ = 1.0 %) with $y$ = 0 (sample A, yellow), $4.4 \times 10^{19}$ cm$^{-3}$ (sample B, orange), and $4.8 \times 10^{20}$ cm$^{-3}$ (sample C, red), and the $Ge_{1-x}Fe_x$ films ($x$ = 2.3 %) with $y$ = 0 (sample D, blue) and $4.4 \times 10^{19}$ cm$^{-3}$ (sample E, green). In the inset of Fig. 3, the same *M-H* curves near zero field (± 300 Oe) are also shown, in which clear hysteresis apparently indicates ferromagnetism in the $Ge_{1-x}Fe_x$ films. The saturation magnetization $M_S$ was estimated by subtracting the diamagnetic signal of the substrate from the raw *M-H* curves, and these are listed in Table I. Note that paramagnetic signals from the $Ge_{1-x}Fe_x$ films were neglected in this procedure. Although $M_S$ of the $Ge_{1-x}Fe_x$ ($x$ = 1.0 %, 2.3 %) films ranges from 1.5 to 0.8 $\mu_B$/Fe atom, there appeared to be no correlation between $M_S$ and $p$. Since it was very difficult to estimate the Curie temperature $T_C$ from the Arrott plots of the *M-H* curves in Fig. 3 (partially due to the subtracting procedure described above), $T_C$ listed in Table I was estimated from the Arrott plots of MCD spectra measured in the temperature range from 5 to 20 K (see supplementary material).[19] The slight difference in $T_C$ between various $p$ is not essential but due to the slight difference in sample quality, because $T_C$ of the $Ge_{1-x}Fe_x$ films significantly depends on the growth temperature as shown in the previous study.[20]

In general, the ferromagnetism in FMSs is considered to be induced by itinerant carriers, and the mean field Zener model[3,21] or double-exchange model[22,23] have been frequently used to explain the observed magnetic properties.[4] In the boron-doped $Ge_{1-x}Fe_x$ films (sample B, C, and E), the resistivity (Fig. 2 (a) and (b)) and mobility (Fig. 2 (e) and (f)) indicate that the Fermi level exists in the valence band. Moreover, the increase in $p$ by two orders of magnitude with increasing $y$ (Fig. 2 (c) and (d)) is attributable to a significant shift of the Fermi level in the valence band of the $Ge_{1-x}Fe_x$ films. However, it was found that there is no correlation between $p$ and the magnetic properties, as shown in Table I and Fig. 3. Even though the transport in the $Ge_{1-x}Fe_x$ ($x$ = 1.0 %, 2.3 %) changed from insulating to metallic, $T_C$ was almost unchanged.  This result presents a contrast to the hole-induced ferromagnetism in (III,Mn)V FMSs.[1,2]  Namely, the mean field Zener model is not applicable, since this model derives a monotonic increase in $T_C$ with $p$.  At present, the double exchange model is one of the possible candidate mechanisms, but we cannot conclude whether this or other models are applicable or not, and further study is needed to clarify the origin of the ferromagnetism in $Ge_{1-x}Fe_x$ films.

In summary, we have revealed the transport properties of $Ge_{1-x}Fe_x$ ($x$ = 1.0 and 2.3 %) with and without boron doping.  By changing the boron doping concentration, we obtained both insulating and metallic $Ge_{1-x}Fe_x$ thin films with the hole concentration $p$ ranging from ~$10^{18}$ to ~$10^{20}$ cm$^{-3}$.  In contrast to the III-V based FMSs, no correlation was observed between $p$ (insulating or metallic) and $T_C$ (magnetic properties).  The present results provide important clues to understanding the origin of ferromagnetism in $Ge_{1-x}Fe_x$.

This work was partly supported by Grant-in-Aids for Scientific Research including Specially Promoted Research, the Project for Developing Innovation Systems of MEXT, and FIRST Program of JSPS. The authors acknowledge the support of Global COE and Outstanding Graduate School Programs "Secure-life Electronics" sponsored by JSPS and MEXT.

**References**

[1] H. Ohno, J. Magn. Magn. Mater. 200, 110 (1999).

[2] M. Tanaka, J. Vac. Sci. Technol. B 16, 2267 (1998).

[3] T. Dietl, H. Ohno, F. Matsukura, J. Cibert, and D. Ferrand, Science 287, 1019 (2000).

[4] N. Samarth, Nature Mater. 11, 360 (2012).

[5] Y. D. Park, A. T. Hanbicki, S. C. Erwin, C. S. Hellberg, J. M. Sullivan, J. E. Mattson, T. F. Ambrose, A. Wilson, G. Spanos, and B. T. Jonker, Science 295, 651 (2002).

[6] M. Jamet, A. Barski, T. Devillers, V. Poydenot, R. Dujardin, P. Bayle-Guillemaud, J. Rothman, E. Bellet-Amalric, A. Marty, J. Cibert, R. Mattana, and S. Tatarenko, Nature Materials 5, 653 (2006).

[7] J.-S. Kang, G. Kim, S. C. Wi, S. S. Lee, S. Choi, S. Cho, S. W. Han, K. H. Kim, H. J. Song, H. J. Shin, A. Sekiyama, S. Kasai, S. Suga, and B. I. Min, Phys. Rev. Lett. 94, 147202 (2005).

[8] S. Sugahara, K. L. Lee, S. Yada, and M. Tanaka, Jpn. J. Appl. Phys. 44, L1426 (2005).

[9] A. P. Li, C. Zeng, K. van Benthem, M. F. Chisholm, J. Shen, S. V. S. Nageswara Rao, S. K. Dixit, L. C. Feldman, A. G. Petukhov, M. Foygel, and H. H. Weitering, Phys. Rev. B 75, 201201(R) (2007).

[10] J. Chen, K. L. Wang, and K. Galatsis, Appl. Phys. Lett. 90, 012501 (2007).

[11] T. Devillers, M. Jamet, A. Barski, V. Poydenot, P. Bayle-Guillemaud, E. Bellet- Amalric, S. Cherifi, and J. Cibert, Phys. Rev. B 76, 205306 (2007).

[12] Y. Wang, F. Xiu, J. Zou, K. L. Wang, and A. P. Jacob, Appl. Phys. Lett. 96, 051905 (2010).

[13] F. Xiu, Y. Wang, J. Kim, A. Hong, J. Tang, A. P. Jacob, J. Zou, and K. L. Wang, Nature Mater. 9, 337 (2010).

[14] S. Yada, R. Okazaki, S. Ohya, and M. Tanaka, Appl. Phys. Express 3, 123002 (2010).

[15] Y. Shuto, M. Tanaka, and S. Sugahara, J. Appl. Phys. 99, 08D516 (2006).

[16] Y. Shuto, M. Tanaka, and S. Sugahara, Appl. Phys. Lett. 90, 132512 (2007).

[17] S. Mirabella, G. Impellizzeri, A. M. Piro, E. Bruno, and M. G. Grimaldi, Appl. Phys. Lett. 92, 251909 (2008).

[18] Y. Shuto, M. Tanaka, and S. Sugahara, Jpn. J. Appl. Phys. 47, 7108 (2008).

[19] See supplementary material at (URL) for the Arrott plots of the MCD signals from the $Ge_{1-x}Fe_x$ films.

[20] Y. Shuto, M. Tanaka, and S. Sugahara, Phys. Stat. Sol. (c) 3, 4110 (2006).

[21] C. Zener, Phys. Rev. 82, 403 (1951).

[22] H. Akai, Phys. Rev. Lett. 81, 3002 (1998).

[23] K. Hirakawa , S. Katsumoto, T. Hayashi, Y. Hashimoto, and Y. Iye, Phys. Rev. B 65, 193312 (2002).

| | $x$ (%) | Boron conc. $y$ (cm$^{-3}$) | Hole conc. $p$ at 300K (cm$^{-3}$) | (004) peak $2\theta$ (deg) | FWHM (deg) | $A$ (Å) | $T_C$ (K) | $M_S$ at 5T ($\mu_B$/Fe atom) |
|---|---|---|---|---|---|---|---|---|
| Ge:B | 0 | 4.4×10$^{19}$ | 2.8×10$^{19}$ | 65.910 | 0.488 | 5.664 | - | - |
| A | 1.0 | Undoped | 1.9×10$^{18}$ | 65.841 | 0.346 | 5.669 | 7 | 1.3 |
| B | 1.0 | 4.4×10$^{19}$ | 2.8×10$^{19}$ | 65.864 | 0.326 | 5.668 | 7 | 1.7 |
| C | 1.0 | 4.8×10$^{20}$ | 2.0×10$^{20}$ | 66.164 | 0.410 | 5.645 | < 5 | 0.8 |
| D | 2.3 | Undoped | 1.2×10$^{18}$ | 65.943 | 0.373 | 5.662 | 15 | 1.2 |
| E | 2.3 | 4.4×10$^{19}$ | 1.7×10$^{19}$ | 65.953 | 0.344 | 5.661 | 17 | 0.9 |

TABLE I. Parameters of the samples of 100-nm-thick Ge$_{1-x}$Fe$_x$ films grown on Si-on-insulator (001) substrates: The Fe content $x$, boron doping concentration $y$, $2\theta$ angle and full width half maximum (FWHM) of the Ge$_{1-x}$Fe$_x$ (004) X-ray diffraction peaks, lattice constant $a$, Curie temperature $T_C$ determined from magnetic circular dichroism (MCD) and Arrott plots, and magnitude of the saturation magnetization $M_S$ per Fe atom estimated by SQUID.

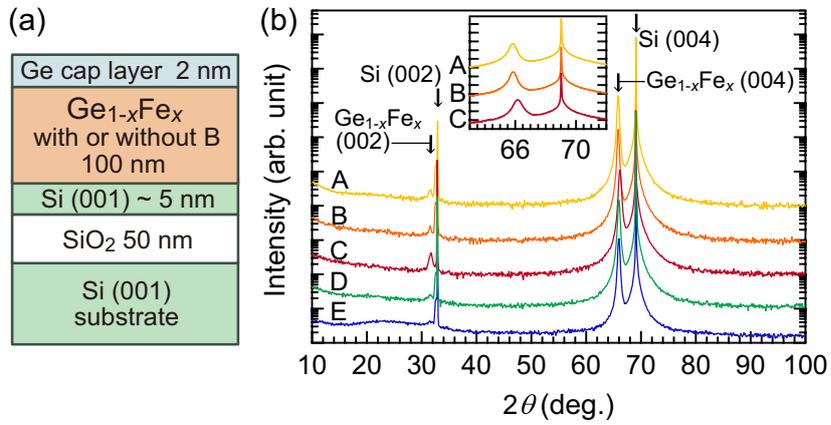

FIG. 1. (a) Schematic structure of the samples consisting of Ge (2 nm) / B-doped $Ge_{1-x}Fe_x$ (100 nm) / undoped Si (~5 nm) / $SiO_2$ (50 nm) / Si(001). (b) X-ray diffraction $\theta$-$2\theta$ spectra of the $Ge_{1-x}Fe_x$ films on silicon-on-insulator (SOI) substrates in the angular range from 10° to 100° in $2\theta$. Inset shows the magnified $Ge_{1-x}Fe_x$ (004) and Si (004) peaks of the $Ge_{0.99}Fe_{0.01}$ films on SOI substrates.

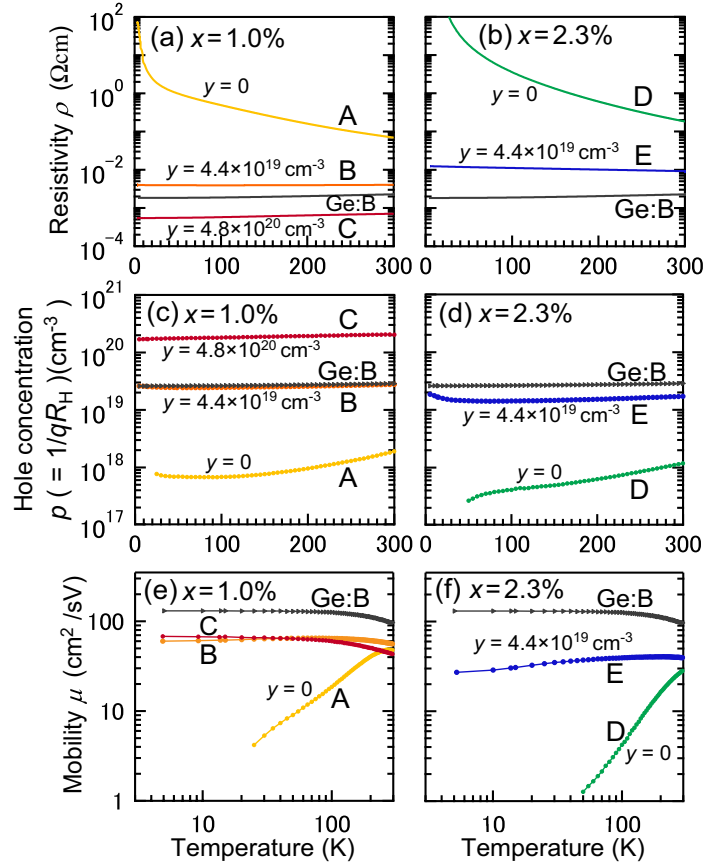

FIG. 2. (a,b) Temperature dependence of the resistivity in the $Ge_{1-x}Fe_x$ films at Fe content (a) $x$ = 1.0 % and (b) 2.3 %, with or without boron doping. The boron doping concentrations $y$ of the $Ge_{1-x}Fe_x$ films ($x$ = 1.0 %) are $y$ = 4.4×10$^{19}$ cm$^{-3}$ (orange), 4.8×10$^{20}$ cm$^{-3}$ (red), and 0 (undoped, yellow), respectively, and those of the $Ge_{1-x}Fe_x$ films ($x$ = 2.3 %) are $y$ = 4.4×10$^{19}$ cm$^{-3}$ (blue), and 0 (undoped, green). The temperature dependence of boron-doped Ge at a concentration of $y$ = 4.4×10$^{19}$ cm$^{-3}$ on SOI (001) substrate is also shown as a reference. (c), (d) Temperature dependence of the hole concentration $p$ in the $Ge_{1-x}Fe_x$ films at Fe content (c) $x$ = 1.0 % and (d) $x$ = 2.3 %, with and without boron doping. (e), (f) Temperature dependence of the hole mobility $\mu$ in the $Ge_{1-x}Fe_x$ films at Fe content (e) $x$ = 1.0 % and (f) 2.3 % with and without boron doping, and $\mu$ in boron-doped ($y$ = 4.4×10$^{19}$ cm$^{-3}$) Ge is also shown as a reference.

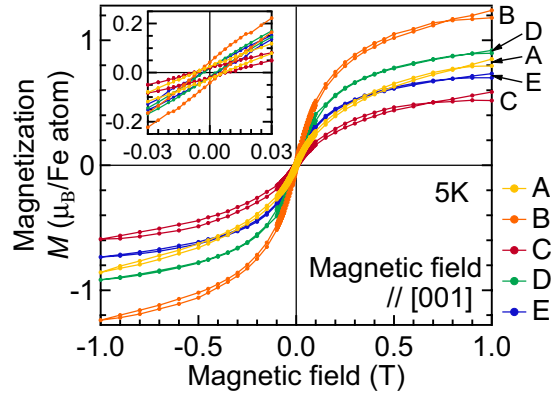

FIG. 3. Magnetic field dependence of the magnetization of the $Ge_{1-x}Fe_x$ films ($x$ = 1.0 %) with boron doping at $y$ = 4.4×10$^{19}$ cm$^{-3}$ (orange), 4.8×10$^{20}$ cm$^{-3}$ (red), and without boron doping (yellow), and the $Ge_{1-x}Fe_x$ films ($x$ = 2.3 %) with $y$ = 4.4×10$^{19}$ cm$^{-3}$ (green) and 0 (blue), respectively. The magnetic field is applied perpendicular to the film plane along the [001] axis. The inset shows the close-up view near zero magnetic field.

**Supplementary material**

**Carrier transport properties of the Group-IV ferromagnetic semiconductor $Ge_{1-x}Fe_x$ with and without boron doping**

Yoshisuke Ban, Yuki Wakabayashi, Ryota Akiyama, Ryosho Nakane and Masaaki Tanaka

*Department of Electronic Engineering and Information Systems, The University of Tokyo,*

*7-3-1 Hongo, Bunkyo-ku, Tokyo 113-8656, Japan*

**Arrott plots of the MCD signals from the $Ge_{1-x}Fe_x$ films**

In order to estimate the Curie temperature $T_C$ of the $Ge_{1-x}Fe_x$ films, we carried out magnetic circular dichroism (MCD) characterizations, because the MCD signal is proportional to the magnetization of the magnetic layer ($Ge_{1-x}Fe_x$) and does not contain the diamagnetic contribution of the substrate. Fig. S1. shows the magnetic field dependence of the reflection MCD intensity and the Arrott plots, $(MCD)^2 - B/MCD$, where MCD is the reflection MCD intensity at a photon energy of local maximum in each MCD spectrum and $B$ is the magnetic field applied perpendicular to the film plane, for the boron-doped $Ge_{1-x}Fe_x$ films ($x = 1.0$ %) with $y = 0$ (Sample A), $y = 4.4 \times 10^{19}$ cm$^{-3}$ (Sample B), $y = 4.8 \times 10^{20}$ cm$^{-3}$ (Sample C), the $Ge_{1-x}Fe_x$ films ($x = 2.3$ %) with $y = 0$ (Sample D), and $y = 4.4 \times 10^{19}$ cm$^{-3}$ (Sample E). The photon energy of local maximum at which the MCD intensity was measured was 1.93 eV for Sample A, 1.93 eV for Sample B, 1.85 eV for Sample C, 1.95 eV for Sample D, and 1.96 eV for Sample E. The $T_C$ values of the $Ge_{1-x}Fe_x$ films were estimated by these Arrott plots and are listed in Table I.

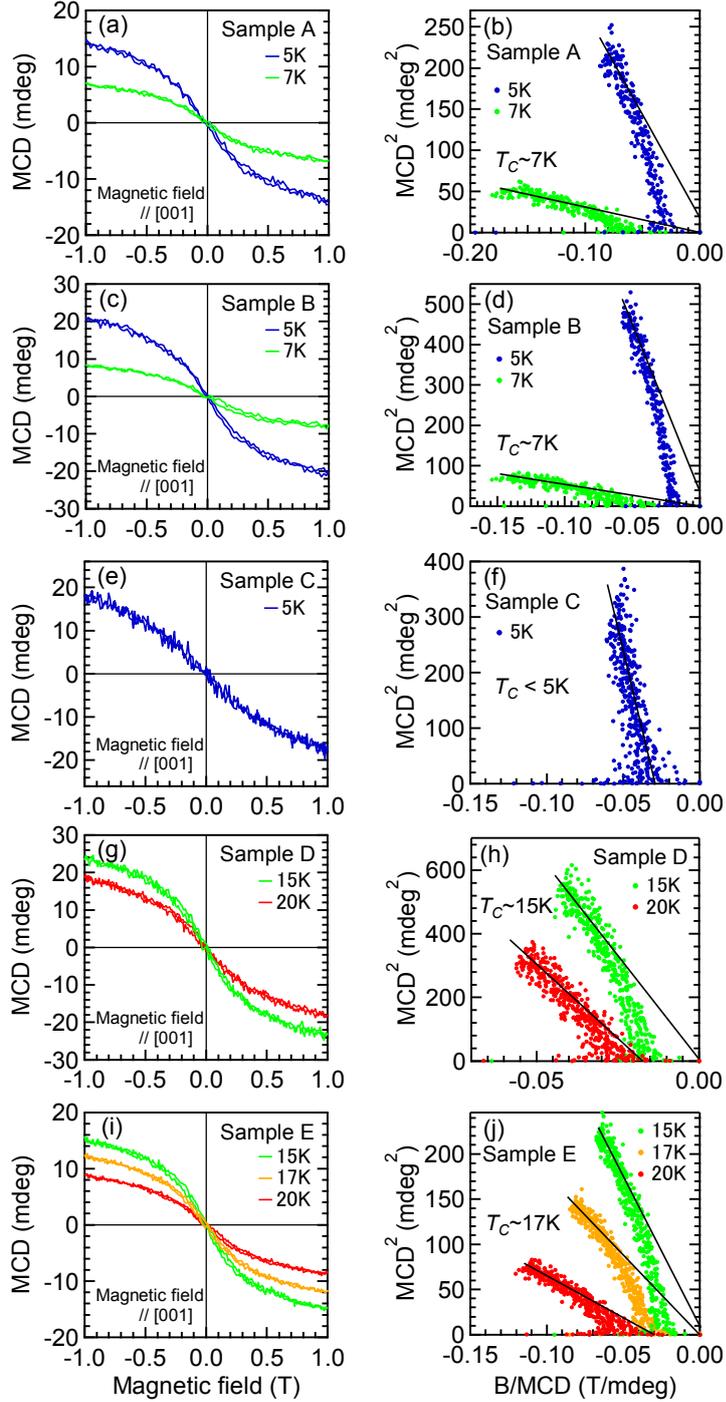

FIG. S1. Magnetic field dependence of the MCD intensity and Arrott plots, (MCD)$^2$ − $B$/MCD, for the boron-doped Ge$_{1-x}$Fe$_x$ films ($x$ = 1.0 %) with (a) (b) $y$ = 0 (Sample A), (c) (d) $y$ = 4.4×10$^{19}$ cm$^{-3}$ (Sample B), (e) (f) $y$ = 4.8×10$^{20}$ cm$^{-3}$ (Sample C), the Ge$_{1-x}$Fe$_x$ films ($x$ = 2.3 %) with (g) (h) $y$ = 0 (Sample D), and (i) (j) $y$ = 4.4×10$^{19}$ cm$^{-3}$ (Sample E), respectively. These MCD signals were measured in the temperature range from 5 to 20 K while a perpendicular magnetic field was swept from -1 to 1 T.

15